\documentclass[aps,pre,twocolumn,10pt]{revtex4-2}

\usepackage[english]{babel}
\usepackage[utf8]{inputenc}
\usepackage[margin=1.in]{geometry}
\usepackage{enumerate}
\usepackage{graphicx}
\usepackage{xcolor}
\usepackage{amsmath,amsfonts,bm,braket}

\usepackage{tikz}
\tikzset{
  common/.style={fill,name=#1,node contents={},inner sep=0,minimum size=3},
  disc/.style={circle,common=#1},
  square/.style={rectangle,common={#1}},
}
\usepackage[export]{adjustbox}

\usepackage{csvsimple}
\usepackage{booktabs}
\usepackage{siunitx}

\usepackage[breaklinks]{hyperref}
\hypersetup{
    colorlinks,
    linkcolor={red!80!black},
    citecolor={green!70!black},
    urlcolor={blue!80!black}
}

\newcommand*{\diff}{\mathop{}\!\mathrm{d}}

\newcommand*{\mmean}[1]{\mathinner{\langle\!\langle {#1} \rangle\!\rangle}}

\DeclareMathOperator{\tr}{tr}
\DeclareMathOperator{\erfc}{erfc}
\DeclareMathOperator{\erf}{erf}
\DeclareMathOperator{\sign}{sign}

\begin{document}

\title{Comment on ``Storage properties of a quantum perceptron''}
\author{Mauro Pastore}
\email{mpastore@ictp.it}
\affiliation{The Abdus Salam International Centre for Theoretical Physics, Strada Costiera 11, 34151 Trieste, Italy}
\date{\today}

\begin{abstract}
The recent paper ``Storage properties of a quantum perceptron'' [Phys. Rev. E 110, 024127] considers a quadratic constraint satisfaction problem, motivated by a quantum version of the perceptron. In particular, it derives its critical capacity, the density of constraints at which there is a satisfiability transition. The same problem was considered before in another context (classification of geometrically structured inputs, see [Phys. Rev. Lett. 125, 120601; Phys. Rev. E 102, 032119; J.~Stat. Mech. (2021) 113301]), but the results on the critical capacity drastically differ. In this note, I substantiate the claim that the derivation performed in the quantum scenario has issues when inspected closely, I report a more principled way to perform it and I evaluate the critical capacity of an alternative constraint satisfaction problem that I consider more relevant for the quantum perceptron rule proposed by the article in question.
\end{abstract}

\maketitle

\section{Introduction}
The recent paper~\cite{gratsea2024} studies the following Gardner volume:
\begin{equation}
V = \int_{\mathbf{w}}\,\prod_{\mu=1}^p \theta\Bigl(\frac{1}{m}|\mathbf{i}^\mu \cdot \mathbf{w}|^2 -\kappa \Bigr) \rho(\mathbf{w})\,,
\label{eq:vol0}
\end{equation}
where $\mathbf{w}$ are the parameters of a (quantum) perceptron and $\mathbf{i}^\mu$ are input vectors with i.i.d. elements, either binary or standard Gaussian,
see Eq.~(6) in the paper. This volume measures the space of configurations of the vector $\mathbf{w}$ that are solutions of the constraint satisfaction problem defined by the set of $p$ constraints enforced by the Heaviside step functions, for a typical realization of the inputs $\{\mathbf{i}^\mu\}$. This volume is equivalent, for $\mathbf{i}^\mu\in \mathbb{R}^m$ (Ref.~\cite{gratsea2024} considers complex inputs in general, as they represent coefficients spanning a Hilbert space; however, as far as I can tell the main running examples -- binary and Gaussian inputs --  are real), to the one introduced in~\cite{rotondo2020,pastore2020,pastore2021} in the context of the unsupervised classification of simplexes by a classical $m$-dimensional perceptron. See, for comparison, Eq.~(37) in \cite{pastore2021}. However, the results on the storage capacity (the critical value of the density of constraints $\alpha=p/m$ at which the problem becomes typically unsatisfiable) drastically differ: \cite{gratsea2024} finds a finite value of $\alpha_c(\kappa)$ for $\kappa\to 0$, while we found a diverging result, $\alpha_c(\kappa\to 0)\to+\infty$. In this comment, I support the claim that the approach of~\cite{gratsea2024} is based on uncontrolled approximations that make their result not grounded. My main concerns are:
\begin{itemize}
    \item The way the critical capacity is obtained in the case of binary inputs is based on approximations that are not valid in the present case (see section~\ref{sec:binary}).
    \item The way the critical capacity is obtained in the case of Gaussian inputs is based on a perturbative approach that, according to the authors themselves, can possibly predict a non-physical behaviour (see section~\ref{sec:Gaussian}); I will discuss a non-perturbative way to do the same calculation in section~\ref{sec:non_perturbative}, that predicts a standard SAT/UNSAT transition in the phase diagram of the model.
    \item The meaning of the volume~\eqref{eq:vol0} is not the one claimed in the paper. I will discuss this issue in section~\ref{sec:meaning}, and propose what I think is a better alternative.
\end{itemize}

Let me first state that in $\kappa = 0$ the storage capacity of $V$ \emph{has} to be infinite: this is indeed the critical value of the density of constraints at which the constraint satisfaction problem defined by the volume passes from being satisfiable to be unsatisfiable, but for $\kappa=0$ all the constraints in Eq.~\eqref{eq:vol0} are trivially satisfied for any $\mathbf{w}$ and any instance of the input patterns, being enforced by step functions of non-negative quantities. However, there could still be a discontinuity that makes $\lim_{\kappa \to 0^+} \alpha_c(\kappa) \neq \alpha_c(0)$. In the following, I will argue why I do not think this is the case.

For illustration purposes, I will specialize my reasoning to the case of spherical weights,
\begin{equation}
	\rho(\mathbf{w}) = \frac{1}{V_{0}}\delta(|\mathbf{w}|^2-m) \,,
\end{equation}
but my concerns remain true even in the case of binary weights. The normalizing constant in the spherical case is
\begin{equation}
V_{0} = \int_{\mathbf{w}} \delta(|\mathbf{w}|^2-m) 
= \frac{2 \pi^{\frac{m}{2}} m^{\frac{m-1}{2}}}{\Gamma(m/2)}  \sim e^{\frac{m}{2}[1+\log(2\pi)]}\,,
\end{equation}
the latter relation ``$\sim$'' meaning ``asymptotically for $m$ large''.

\section{Two representations of the Heaviside step function}

 As the derivations of~\cite{gratsea2024} and~\cite{rotondo2020,pastore2020,pastore2021} are performed in different ways, let me reproduce here in some detail the steps performed by the first reference, and how they differ from ours. The calculation is based on the replica method. The replicated volume is
\begin{equation}
V^n =\int_{\{\mathbf{w}^a\}} \prod_{a=1}^n\rho(\mathbf{w}^a) \prod_{\mu=1}^p \theta \Bigl(\frac{1}{m}|\mathbf{i}^\mu \cdot \mathbf{w}^a|^2 -\kappa \Bigr).
\end{equation}
Introducing a delta function and using its Fourier representation, Ref.~\cite{gratsea2024} writes (see Eq.~(9) and (B2))
\begin{equation}
\begin{aligned}
&\theta\Bigl(\frac{1}{m}|\mathbf{i}^\mu \cdot \mathbf{w}^a|^2 -\kappa \Bigr) \\
&\quad= \int \diff \lambda^a_\mu \,\theta(\lambda^a_\mu - \kappa)\delta\Bigl(\lambda^a_\mu - \frac{|\mathbf{i}^\mu \cdot \mathbf{w}^a|^2}{m} \Bigr) \\
&\quad=  \int \frac{\diff \lambda^a_\mu \diff x^a_\mu}{2\pi} \, \theta(\lambda^a_\mu - \kappa) e^{i x^a_\mu \lambda^a_\mu - i x^a_\mu\frac{|\mathbf{i}^\mu \cdot \mathbf{w}^a|^2}{m}}  \,.
\end{aligned}
\label{eq:thetaGratsea}
\end{equation}
This step is performed differently in~\cite{rotondo2020,pastore2020,pastore2021}:
\begin{equation}
\begin{aligned}
&\theta \Bigl(\frac{1}{m}|\mathbf{i}^\mu \cdot \mathbf{w}^a|^2 -\kappa \Bigr) 
= 
\theta \Bigl(\frac{1}{m}(\mathbf{i}^\mu \cdot \mathbf{w}^a)^2 -\kappa \Bigr)\\
&= \int \diff \lambda^a_\mu \,\theta[(\lambda^a_\mu)^2 - \kappa]\delta \Bigl(\lambda^a_\mu - \frac{\mathbf{i}^\mu \cdot \mathbf{w}^a}{\sqrt{m}} \Bigr)\\
&=\int \frac{\diff \lambda^a_\mu \diff x^a_\mu}{2\pi} [\theta(\lambda^a_\mu - \sqrt{\kappa})+\theta(-\lambda^a_\mu - \sqrt{\kappa}) ]\\
&\hspace{12em} \times e^{i x^a_\mu \lambda^a_\mu - i x^a_\mu \frac{\mathbf{i}^\mu \cdot \mathbf{w}^a}{\sqrt{m}} }
,
\end{aligned}
\label{eq:thetaMine}
\end{equation}
the first step being trivially true as $|\lambda|^2 = \lambda^2$ for any $\lambda\in \mathbb{R}$, and the last step being true because the constraint $\lambda^2 -\kappa >0$ implies $\lambda<-\sqrt{\kappa} \vee \lambda>\sqrt{\kappa}$.

Both Eq.~\eqref{eq:thetaGratsea} and~\eqref{eq:thetaMine} seem to me legitimate formal way to write the step function, so the final result of the calculation cannot be different due to this choice. From these representations, the calculation proceeds integrating over the input distribution, that is to evaluate the quantity
\begin{equation}
\begin{aligned}
    G_1 
    &= \frac{1}{p} \log \mmean{
\prod_{\mu=1}^p\prod_a \theta \Bigl(\frac{1}{m}|\mathbf{i}^\mu \cdot \mathbf{w}_a|^2 -\kappa \Bigr)}\\
&= \log \mmean{
\prod_a \theta \Bigl(\frac{1}{m}|\mathbf{i} \cdot \mathbf{w}_a|^2 -\kappa \Bigr)}\,,
\end{aligned}
\label{eq:G1_def}
\end{equation}
where the average over the inputs $\mmean{\cdot}$ is factorized over the index $\mu = 1,\cdots, p$.
Let me first consider the approach of~\cite{gratsea2024} based on Eq.~\eqref{eq:thetaGratsea}, splitting the discussion in the two cases considered there for the input distribution, while postponing the discussion on the results obtained from the representation~\eqref{eq:thetaMine} to section~\ref{sec:non_perturbative}.

\section{Binary inputs\label{sec:binary}}

For binary inputs, the average~\eqref{eq:G1_def} using the representation~\eqref{eq:thetaGratsea} gives
\begin{equation}
\begin{aligned}
    G_1 ={}& \log \int \left[\prod_a  \frac{\diff \lambda^a \diff x^a}{2\pi} \, \theta(\lambda^a - \kappa)\right] e^{i \sum_a x^a \lambda^a}\\
    &\times \mmean {e^{ - i \sum_a x^a \frac{1}{m}\sum_{k,\ell} i_k i_\ell  w^a_k w^a_\ell}}\\
    ={}&\log \int \left[\prod_a  \frac{\diff \lambda^a \diff x^a}{2\pi} \, \theta(\lambda^a - \kappa)\right] e^{i \sum_a x^a \lambda^a}\\
    &\times e^{- i \sum_a x^a\sum_{k} \frac{(w^a_k)^2}{m} } \mmean { \prod_{k\neq\ell} e^{ - i \sum_a x^a  i_k i_\ell   \frac{w^a_k w^a_\ell}{m}}}.
\end{aligned}
\label{eq:G1_binary1}
\end{equation}
Then, Ref.~\cite{gratsea2024} Eq.~(29) writes
\begin{multline}
     \mmean { \prod_{k\neq\ell} e^{ - i \sum_a x^a  i_k i_\ell   \frac{w^a_k w^a_\ell}{m}}}
     \\
     \approx \prod _{k\neq\ell} \cos \Bigl( \frac{2}{m} \sum_a x^a w^a_k w^a_\ell \Bigr),
    \label{eq:approx1}
\end{multline}
which, using the expansion $\log \cos(x) \approx -x^2/2$, yields Eq.~(31)
\begin{equation}
\begin{aligned}
    G_1 &= \log \int \left[\prod_a  \frac{\diff \lambda^a \diff x^a}{2\pi} \, \theta(\lambda^a - \kappa)\right] \\
    &\times e^{i \sum_a x^a (\lambda^a -1)  - \sum_a (x^a)^2 - 2 \sum_{a<b} x^a x^b q^{ab}}\,,
\end{aligned}
    \label{eq:G1_binary2}
\end{equation}
where $q^{ab} = \sum_k w_k^a w_k^b /m$.
I claim that \emph{the steps to pass from Eq.~\eqref{eq:G1_binary1} to~\eqref{eq:G1_binary2} are not justified in the present context}. To see this, let me consider the series expansion in $1/m$ of the term inside the angle brackets in~\eqref{eq:G1_binary1}:
\begin{equation}
\begin{aligned}
&\mmean{\prod_{k\neq\ell} e^{ -  \frac{i}{m} \sum_a x^a i_k i_\ell w^a_k w^a_\ell}   } \\
&=
 1 - \frac{1}{2 m^2} \sum_{a,b} x^a x^b \sum_{k\neq\ell,r\neq s} \mmean{i_k i_\ell i_r i_s} w^a_k w^a_\ell w^b_r w^b_s \\
& +  \frac{i}{6 m^3} \sum_{a,b,c} x^a x^b x^c \\
&\quad \times\sum_{k\neq\ell,r\neq s,t\neq u} \mmean{i_k i_\ell i_r i_s i_t i_u} w^a_k w^a_\ell w^b_r w^b_s w^c_t w^c_u \\
& +  \frac{1}{24 m^4} \sum_{a,b,c,d} x^a x^b x^c x^d \\
&\quad \times\!\sum_{\substack{k\neq\ell,r\neq s\\t\neq u,v\neq z}} \mmean{i_k i_\ell i_r i_s i_t i_u i_v i_z} w^a_k w^a_\ell w^b_r w^b_s w^c_t w^c_u w^d_v w^d_z  \\
&+ \cdots
\end{aligned}
\end{equation}
The expectations over the inputs can be done by noticing that any odd contraction of indices gives 0, so
\begin{equation}
\begin{aligned}
&\mmean{\prod_{k\neq\ell} e^{ -  \frac{i}{m} \sum_a x^a i_k i_\ell w^a_k w^a_\ell}   } \\
&=
 1 - \frac{1}{m^2} \sum_{a,b} x^a x^b \sideset{}{'}\sum_{k,\ell} w^a_k w^b_k w^a_\ell  w^b_\ell\\
 & +  \frac{4i}{3 m^3} \sum_{a,b,c} x^a x^b x^c \sideset{}{'}\sum_{k,\ell, r}  w^a_k w^b_k w^a_\ell w^c_\ell  w^b_r w^c_r \\
 & +  \frac{1}{2m^4} \sum_{a,b,c,d} x^a x^b x^c x^d  \sideset{}{'}\sum_{k,\ell,r,s} w^a_k w^b_k w^a_\ell  w^b_\ell w^c_r w^d_r w^c_s  w^d_s  \\
 & +  \frac{2}{m^4}  \sum_{a,b,c,d} x^a x^b x^c x^d\sideset{}{'}\sum_{k,\ell,r, s} w^a_k w^b_k w^a_\ell w^c_\ell  w^b_r w^d_r  w^c_s  w^d_s \\
 &+ \frac{1}{12 m^4}  \sum_{a,b,c,d} x^a x^b x^c x^d \sideset{}{'}\sum_{k,\ell} w_k^a w_k^b w_k^c w_k^d w_\ell^a w_\ell^b w_\ell^c w_\ell^d 
 \\
 &+
 \cdots\\
\end{aligned}
\end{equation}
The numerical factors come from the combinatorics of index contraction, and $\sum'$ denotes a sum over all different indices. One can write everything in terms of the overlap $q^{ab}$ by noticing that, for large $m$, (i)~we can drop the last term, as it has only two summation indices but a factor $1/m^4$ suppressing it; (ii) we can relax the constraints on the sums $\sum' \to \sum$, as the difference between the two are again terms with less summation indices that become negligible. I obtain:
\begin{equation}
\begin{aligned}
&\mmean{\prod_{k\neq\ell} e^{ -  \frac{i}{m} \sum_a x_a i_k i_\ell w^a_k w^a_\ell}   } \approx 
 1 - \sum_{a,b} x^a x^b (q^{ab})^2 \\
 &+  \frac{4i}{3} \sum_{a,b,c} x^a q^{ab} x^b q^{bc} x^c q^{ca} + \frac{1}{2} \Bigl[\sum_{a,b} x^a x^b (q^{ab})^2  \Bigr]^2 \\
 &+  2 \sum_{a,b,c,d} x^a q^{ab} x^b q^{bc} x^c q^{cd} x^d q^{da} + \cdots
\end{aligned}
\end{equation}
All these terms are finite for large $m$, even in the case of small off-diagonal $q$ (due to the contributions at $a=b$), and so the truncation of this series (taking only the terms corresponding to the series of the cosine, while discarding the rest) is never justified.

\section{Gaussian inputs\label{sec:Gaussian}}
For (real) Gaussian input, the average gives
\begin{multline}
G_1= \log \int \Bigl[\prod_a \frac{\diff \lambda^a \diff x^a}{2\pi}  \theta(\lambda^a - \kappa)\Bigr] \\
\times e^{i \sum_a x^a \lambda^a - \frac{1}{2}\log \det(\mathbb{I}_m + \frac{2i}{m} \sum_a x^a \mathbf{w}^a\mathbf{w}^{a\top})} \,.
\end{multline}
Discrepancies w.r.t.~\cite{gratsea2024}, Eq.~(21), are possibly due to minor typos. Then, assuming replica symmetry $q^{ab} = (1-q)\delta_{ab}+q$, Ref.~\cite{gratsea2024} finds a self-consistent equation for the determinant in this equation depending on $q$, solves it for small $q$ and ``continues'' the perturbative result to any value of $q$, finding a peculiar easy/hard phase transition, different in nature w.r.t. the standard SAT/UNSAT transition of the classical Gardner case. The authors however comment: ``This behavior might be the result of expansion in $q$ that we used to obtain the effective potential''. I will show in the next section how a non-perturbative approach does not exhibit any peculiar phenomenology of this kind.

Here, I simply notice that, expanding the log-determinant (or trace-log), one gets
\begin{multline}
-\frac{1}{2} \tr \log\Bigl(\mathbb{I} + \frac{2i}{m} A\Bigr)=-\frac{i}{m}\tr(A)-\frac{1}{m^2}\tr(A^2)\\
+\frac{4 i}{3 m^3}\tr(A^3)+\frac{2}{m^4}\tr(A^4)+\cdots\,,
\end{multline}
where I called $A = \sum_a x_a \mathbf{w}^a \mathbf{w}^{a\top}$. All these traces can be expressed straightforwardly in terms of overlaps, as
\begin{equation}
    \begin{aligned}
\frac{\tr(A)}{m} &= \sum_a x^a \,, \\
\frac{\tr(A^2)}{m^2} &= \sum_{a,b} x^a x^b (q_{ab})^2 \\
\frac{\tr(A^3)}{m^3} &= \sum_{a,b,c} x^a q^{ab} x^b  q^{bc} x^c q^{ca} \,,\\
\frac{\tr(A^4)}{m^4} &= \sum_{a,b,c,d} x^a q^{ab} x^b q^{bc} x^c q^{cd} x^d q^{da}
\end{aligned}
\end{equation}
and so on. \emph{This is the same series obtained for binary inputs}: indeed, the assumptions that, for $m$ large, $(1/m^\ell)\sum'_{k_1,\cdots,k_\ell} \to (1/m^\ell)\sum_{k_1,\cdots,k_\ell}$ and $(1/m^\ell)\sum'_{k_1,\cdots,k_{\ell'}} \to 0$ for $\ell'<\ell$, that I made in the previous section, are equivalent to assuming \emph{Gaussian universality of the input distribution}, a property that is routinely assumed based on the above diagrammatics. For this reason, I think that the same technique used in~\cite{gratsea2024} to investigate the case of Gaussian inputs should be applied also to the case of binary inputs, the alternative procedure being doomed by the problem I presented in the previous section.

\section{Non-perturbative approach}\label{sec:non_perturbative}
\begin{figure}
    \centering
    \includegraphics[width=0.9\linewidth]{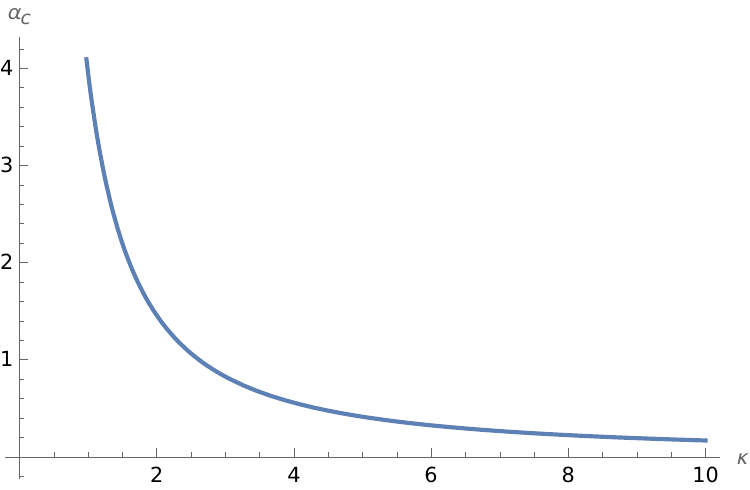}
    \caption{Critical capacity $\alpha_c$ of the volume $V$ as a function of $\kappa$, from Eq.~\eqref{eq:alpha}. Compare with~\cite{gratsea2024}, Fig.~2.}
    \label{fig:alpha}
\end{figure}

Let me now show how one can evaluate non-perturbatively in $q$ the potential $G_1$ using the representation~\eqref{eq:thetaMine}, as we did in \cite{rotondo2020,pastore2020,pastore2021}. In this case,
\begin{align}
&G_1 = \log\! \int\! \Bigl( \prod_a \tfrac{\diff \lambda^a \diff x^a}{2\pi} [\theta(\lambda^a \!- \sqrt{\kappa})+\theta(-\lambda^a\! - \sqrt{\kappa}) ] \Bigr) \nonumber\\
   &\hspace{6em} \times e^{i \sum_a x^a \lambda^a} \mmean{e^{ - i \sum_a x^a \frac{\mathbf{i} \cdot \mathbf{w}^a}{\sqrt{m}} }}\,.
\end{align}
The expectation is factorized over the indices spanning $\mathbb{R}^m$ and gives, both for Gaussian and binary inputs (for the latter case, one can expand in $1/\sqrt{m}$ and observe that the non-Gaussian contributions are suppressed, again a consequence of Gaussian universality of the input distribution),
\begin{align}
&G_1 = \log\! \int\! \Bigl( \prod_a \tfrac{\diff \lambda^a \diff x^a}{2\pi} [\theta(\lambda^a \!- \sqrt{\kappa})+\theta(-\lambda^a\! - \sqrt{\kappa}) ] \Bigr) \nonumber\\
    &\hspace{7em} \times e^{i \sum_a x^a \lambda^a - \frac{1}{2} \sum_{a,b} x^a x^b q^{ab}}\,.
\end{align}
With replica symmetry,
\begin{widetext}
\begin{equation}
\begin{aligned}
    G_1 &= \log \int \Bigl( \prod_a \frac{\diff \lambda^a \diff x^a}{2\pi} [\theta(\lambda^a - \sqrt{\kappa})+\theta(-\lambda^a - \sqrt{\kappa}) ] \Bigr) e^{i \sum_a x^a \lambda^a - \frac{1-q}{2} \sum_{a} (x^a)^2  - \frac{q}{2} \sum_{a,b} x^a x^b } \\
    &=\log \int \frac{\diff y \,e^{-\frac{y^2}{2}}}{\sqrt{2\pi}} \Bigl(\int \frac{\diff \lambda \diff x}{2\pi} [\theta(\lambda - \sqrt{\kappa})+\theta(-\lambda - \sqrt{\kappa}) ]  e^{ - \frac{1-q}{2}  x^2  + i (\lambda+y\sqrt{q} ) x }\Bigr)^n\\
    &=\log \int \frac{\diff y\, e^{-\frac{y^2}{2}}}{\sqrt{2\pi}} \Bigl( \int \frac{\diff \lambda }{\sqrt{2\pi(1-q)}} [\theta(\lambda - \sqrt{\kappa})+\theta(-\lambda - \sqrt{\kappa}) ]  e^{ - \frac{(\lambda+y\sqrt{q} )^2}{2(1-q)}  }\Bigr)^n\\
    &=\log \int \frac{\diff y \,e^{-\frac{y^2}{2}}}{\sqrt{2\pi}} \Bigl[\tfrac{1}{2}\erfc  \Bigl(\tfrac{\sqrt{\kappa} + y\sqrt{q}}{\sqrt{2(1-q)}} \Bigr) + \tfrac{1}{2}\erfc  \Bigl(\tfrac{\sqrt{\kappa} - y\sqrt{q}}{\sqrt{2(1-q)}} \Bigr) \Bigr]^n\\
    &\approx n \int \frac{\diff y \,e^{-\frac{y^2}{2}}}{\sqrt{2\pi}} \log\Bigl[\tfrac{1}{2}\erfc  \Bigl(\tfrac{\sqrt{\kappa} + y\sqrt{q}}{\sqrt{2(1-q)}} \Bigr) + \tfrac{1}{2}\erfc  \Bigl(\tfrac{\sqrt{\kappa} - y\sqrt{q}}{\sqrt{2(1-q)}} \Bigr) \Bigr]\,,
    \label{eq:G1_mine}
\end{aligned}
\end{equation}
\end{widetext}
the last equality being true for small number of replicas $n$ (see Eq.~(39) in \cite{pastore2021}).
From this equation, in the limit $q\to 1$ (at which the volume of solutions shrinks to 0 and different replicas become more and more correlated) the expression for $\alpha_c$ follows (see Eq.~(112) in~\cite{pastore2020}):
\begin{equation}
    \alpha_c(\kappa) = \frac{1}{2} \Bigl[\int_0^{\sqrt{\kappa}} \frac{\diff y\, e^{-\frac{y^2}{2}}}{\sqrt{2\pi}} (\sqrt{\kappa} - y)^2 \Bigr]^{-1}.
    \label{eq:alpha}
\end{equation}
The behaviour of $\alpha_c(\kappa)$ is reported in Fig.~\ref{fig:alpha}. At this value, a classical SAT/UNSAT transition, as the ones described by Gardner, occurs. In this case, however, $\alpha_c(0) = +\infty $, consistently with the observation that the problem is always SAT for $\kappa=0$. Notice however that the problem exhibits a replica symmetry breaking transition in the SAT phase, as shown in~\cite{pastore2021}, so the SAT/UNSAT curve $\alpha_c(\kappa)$ predicted from the RS ansatz is only an approximation of the true one.

I can compare here the perturbative expansion around $q=0$ of the potential $G_1$ in Eq.~\eqref{eq:G1_mine} with the one reported in~\cite{gratsea2024}. From~\eqref{eq:G1_mine} I get
\begin{equation}
    G_1 \approx  n\log[ \erfc( \sqrt{\kappa/2})]- n q^2 \frac{\kappa e^{-\kappa} }{
  2 \pi \erfc(\sqrt{\kappa/2})^2}\,,
\end{equation}
while from~\cite{gratsea2024}, Eq.~(58)
\begin{equation}
    G_1 \approx -\frac{\kappa n}{2} - \frac{1}{2}  n q^2 (2+\kappa)^2 \,.
\end{equation}
Notice that, despite the two expressions being different, it is true that
\begin{equation}
\begin{aligned}
    \lim_{\kappa \to \infty} \frac{\log[ \erfc( \sqrt{\kappa/2})]}{-\frac{\kappa }{2}} &= 1\,, \\ 
    \lim_{\kappa \to \infty} \frac{-\frac{\kappa e^{-\kappa} }{
  2 \pi \erfc(\sqrt{\kappa/2})^2}}{ - \frac{1}{2}  (2+\kappa)^2} &= \frac{1}{2}\,.
\end{aligned}
\end{equation}
The matching (apart for irrelevant numerical factors) of the large-$\kappa$ limits suggests that the perturbative method devised in~\cite{gratsea2024} works for $q\ll \kappa$, rather than $q$ small at given $\kappa$. This observation is only a guess, for the purpose of this comment.

\section{Meaning of the Gardner volume\label{sec:meaning}}

In the last section I established the technical way I consider more principled to evaluate the volume~\eqref{eq:vol0}, as it is (i)~non-perturbative in $q$, as opposed to the perturbative method used in~\cite{gratsea2024} for the case of Gaussian inputs; (ii)~not based on an arbitrary truncation of a series whose terms are all finite, used in~\cite{gratsea2024} for the case of binary inputs (assuming universality over the input distribution, the latter should be treated in the same way as the former). Let me now discuss my most serious concern, regarding the meaning that Ref.~\cite{gratsea2024} would like to give to the volume~\eqref{eq:vol0}. The paper (Eq.~(2)) considers a feedforward neural network with an input-output map of the kind
\begin{equation}
    \sigma^\mu = \theta(|\mathbf{i}^\mu \cdot \mathbf{w}|^2/m - \kappa)\,, \qquad \mu=1,\cdots,p
\end{equation}
where $\theta$ is the activation function, $\kappa$ a positive threshold and the outputs $\sigma^\mu \in \{ 0,1\}$. This input-output map is of course equivalent to
\begin{equation}
    {\sigma^\mu}' = \sign(|\mathbf{i}^\mu \cdot \mathbf{w}|^2/m - \kappa)\,, \qquad \mu=1,\cdots,p
    \label{eq:constraints1}
\end{equation}
where ${\sigma^\mu}' = 2 \sigma^\mu - 1 \in \{\pm 1\}$. The Gardner's program~\cite{gardner1988} on the storage capacity (as opposed to the teacher-student scenario, also considered by Gardner in~\cite{gardner1989}) consists in evaluating the volume of parameters $\mathbf{w}$ able to satisfy the constraints~\eqref{eq:constraints1} \emph{for a random assignment of the labels ${\sigma^\mu}'$}. The constraints can be re-written as
\begin{equation}
    {\sigma^\mu}'(|\mathbf{i}^\mu \cdot \mathbf{w}|^2/m - \kappa)\ge 0 \,, \qquad \mu=1,\cdots,p
    \label{eq:constraints2}
\end{equation}
and the set of parameters can be made more \emph{stable} (robust over input noise) by requiring the stronger 
\begin{equation}
    {\sigma^\mu}'(|\mathbf{i}^\mu \cdot \mathbf{w}|^2/m - \kappa)\ge \kappa' \,, \qquad \mu=1,\cdots,p
    \label{eq:constraints3}
\end{equation}
for some $\kappa'>0$. Notice the different role played by the \emph{threshold} (or bias) $\kappa$, that enters the decision rule, and the \emph{margin} $\kappa'$, that has been added to count only ``stable enough'' configurations of parameters. Given this set of constraints, the Gardner volume should be
\begin{equation}
\label{eq:V'}
V' = \int_{\mathbf{w}}\,\prod_{\mu=1}^p \theta \Bigl[{\sigma^\mu}'\Bigl(\tfrac{1}{m}|\mathbf{i}^\mu \cdot \mathbf{w}|^2 - \kappa\Bigr) - \kappa'\Bigr] \rho(\mathbf{w})\,,
\end{equation}
for a typical instance of the inputs $\mathbf{i}^\mu$ and the random labels ${\sigma^\mu}'$,
rather than~\eqref{eq:vol0}, to which it does not reduce neither for $\kappa$ or $\kappa'\to 0$. For example, consider the case where both $\kappa=\kappa' = 0 $: given that the labels are random, typically half of them will be negative; this means that half of the constraints in $V'$ (on average) are not satisfied, the ones corresponding to ${\sigma^\mu}' = -1$, as $|\mathbf{i}^\mu \cdot \mathbf{w}|^2$ cannot be negative. As a result, the critical capacity in this case is 0 (the problem is UNSAT at any finite value of $p/m$ for $p$, $m$ large). In the next paragraph I report the calculation needed to obtain the critical capacity of the volume $V'$.

\subsection{Capacity of the volume \texorpdfstring{$V'$}{V'}}
\begin{figure}
    \centering
    \includegraphics[width=0.9\linewidth]{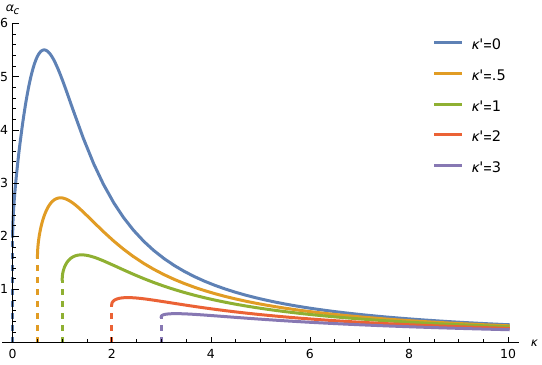}
    \caption{Critical capacity $\alpha_c'$ of the volume $V'$, as a function of $\kappa$, $\kappa'$, from Eq.~\eqref{eq:alpha2}.}
    \label{fig:alpha2}
\end{figure}

Using a representation similar to~\eqref{eq:thetaMine}, the volume $V'$ in Eq.~\eqref{eq:V'} yields a potential
\begin{equation}
    G_1' \approx  n \int \frac{\diff y\, e^{-\frac{y^2}{2}}}{\sqrt{2\pi}} \mathbb{E}_\sigma \log[L_\sigma (y)]\,,
\end{equation}
where $\sigma\sim (\delta_{+1}+ \delta_{-1} )/2$ and
\begin{equation}
    L_\sigma (y) = \int \frac{\diff \lambda\,e^{-\frac{(\lambda+y\sqrt{q})^2}{2(1-q)}}}{\sqrt{2\pi(1-q)}} \theta(\sigma\lambda^2 -\sigma \kappa - \kappa').
\end{equation}
The step function is now enforcing the integration domain $\lambda<-\sqrt{\kappa_+} \vee \lambda>\sqrt{\kappa_+}$ for $\sigma=+$, or $-\sqrt{\kappa_-} <\lambda < \sqrt{\kappa_-}$ for $\sigma = -$ (this second constraint is active only when $\kappa_- > 0$, otherwise $L_-(y) = 0$), where $\kappa_\pm := \kappa \pm \kappa'$. The functions $L_\sigma$ are given more explicitly as
\begin{equation}
    \begin{aligned}
        L_+ &= \tfrac{1}{2}\erfc  \Bigl(\tfrac{\sqrt{\kappa_+} + y\sqrt{q}}{\sqrt{2(1-q)}} \Bigr) + \tfrac{1}{2}\erfc  \Bigl(\tfrac{\sqrt{\kappa_+} - y\sqrt{q}}{\sqrt{2(1-q)}} \Bigr),\\
        L_- &= \theta(\kappa_-)\Bigl[\tfrac{1}{2} \erf \Bigl(\tfrac{\sqrt{\kappa_-} + y\sqrt{q}}{\sqrt{2(1-q)}} \Bigr)+\tfrac{1}{2} \erf \Bigl(\tfrac{\sqrt{\kappa_-} - y\sqrt{q}}{\sqrt{2(1-q)}} \Bigr) \Bigr],
    \end{aligned}
\end{equation}
and admit the $q\to 1$ asymptotics (if $\kappa_->0$)
\begin{equation}
\begin{aligned}
    \log(L_+) &\sim \begin{cases}
     0  & \text{if $y<-\sqrt{\kappa_+}\vee y>\sqrt{\kappa_+}$ }\\
      - \frac{(\sqrt{\kappa_+} + y)^2}{2(1-q)} & \text{if $y\in (-\sqrt{\kappa_+},0)$}\\
      - \frac{(\sqrt{\kappa_+} - y)^2}{2(1-q)} & \text{if $y\in (0,\sqrt{\kappa_+})$}
    \end{cases}\\
    \log(L_-) &\sim \begin{cases}
     - \frac{(\sqrt{\kappa_-} + y)^2}{2(1-q)}  & \text{if $y<-\sqrt{\kappa_-}$}\\
     0 & \text{if $y\in (-\sqrt{\kappa_-},\sqrt{\kappa_-})$}\\
     - \frac{(\sqrt{\kappa_-} - y)^2}{2(1-q)} & \text{if $y>\sqrt{\kappa_-}$}
    \end{cases}
\end{aligned}
\end{equation}
Proceeding as before, one gets the critical capacity
\begin{multline}
    \alpha'_c(\kappa_+,\kappa_-) = \frac{1}{2} \Bigl[\int_0^{\sqrt{\kappa_+}} \frac{\diff y\, e^{-\frac{y^2}{2}}}{\sqrt{2\pi}} \frac{(\sqrt{\kappa_+} - y)^2}{2} \\
    +\int_{\sqrt{\kappa_-}}^{+\infty} \frac{\diff y\, e^{-\frac{y^2}{2}}}{\sqrt{2\pi}} \frac{(\sqrt{\kappa_-} - y)^2}{2}\Bigr]^{-1}\,,
    \label{eq:alpha2}
\end{multline}
which is plotted in Fig.~\ref{fig:alpha2}. Notice that, for $\kappa' = 0$ ($\kappa_+ = \kappa_- = \kappa$), the critical capacity is equal to the Gardner value at $\kappa \to 0$, $\alpha_c'(0,0) = 2$. However, the curve $\alpha_c'(\kappa,\kappa)$ arrives in $\kappa=0$ with a vertical slope, suggesting that the problem is always UNSAT on the $\kappa=0$ axis, in accordance with the preliminary observation made in the above section. As clear from the figure, at given margin $\kappa'$ the curve $\alpha_c'(\kappa+\kappa',\kappa-\kappa')$ is non-monotonic, reaching its maximum at a value $\kappa = \kappa_{\rm max}(\kappa')$ of the threshold.

\bibliographystyle{apsrev4-2}
\bibliography{refs}

\end{document}